\documentstyle[twocolumn,pre,epsfig,aps]{revtex}
\begin{document}
%
\twocolumn[
\hsize\textwidth\columnwidth\hsize\csname@twocolumnfalse\endcsname
%
\title{Is Tsallis thermodynamics nonextensive?}
\date{\today}
\author{Eduard Vives and Antoni Planes}
\address{Departament  d'Estructura  i  Constituents de  la  Mat\`eria,
Universitat de Barcelona, \\ Diagonal 647, 08028 Barcelona, Catalonia
(Spain).}
\maketitle
\begin{abstract}
Within  the  Tsallis  thermodynamics'  framework,  and  using  scaling
properties  of  the  entropy,   we  derive  a  generalization  of  the
Gibbs-Duhem  equation.   The  analysis  suggests a  transformation  of
variables  that  allows   standard  thermodynamics  to  be  recovered.
Moreover, we also generalize Einstein's formula for the probability of
a fluctuation  to occur  by means of  the maximum  statistical entropy
method.   The use  of the  proposed transformation  of  variables also
shows  that fluctuations within  Tsallis statistics  can be  mapped to
those of standard statistical mechanics.
\end{abstract}
]

During the last  few years there has been a great  deal of interest in
studying                  nonextensive                  thermodynamics
\cite{Salazar99,Rajagopal99,Yamano00}.    This    results   from   the
assumption  of  non-additive  statistical  entropies and  the  maximum
statistical  entropy  principle,   following  the  information  theory
formulation    of   statistical    mechanics   proposed    by   Jaynes
\cite{Jaynes57}.  Indeed, besides its relevance in many nonequilibrium
problems,  non-extensivity is  of  interest for  systems of  particles
which show up long-range  interactions \cite{Cannas96}, as occurs in a
number of  ferroic materials \cite{Wadhawam00}  such as ferromagnetic,
ferroelastic,   ferroelectric   solids,   and  astrophysical   systems
\cite{Torres97}.  Within  this framework, Tsallis  statistical entropy
\cite{Tsallis88} has proven to be the only non-additive generalization
of Gibbs-Shannon entropy which satisfies the following properties: (i)
positivity  (it takes  zero value  for absolute  certainty) increasing
monotonously   with  increasing   uncertainty,  and   (ii)  concavity.
However,  many fundamental features  regarding the  connection between
the  formulation of  statistical mechanics  and  thermodynamics remain
unclear.   For instance,  the identification  of  adequate generalized
thermodynamic forces  and the computation  of statistical fluctuations
are still  controversial \cite{Tsallis98}.  In this  letter we clarify
such problems and provide  robust arguments showing the equivalence of
the present formulations  of non-extensive Tsallis thermodynamics with
the standard extensive equilibrium formulation.

Within the Tsallis formalism,  the lack of information associated with
any  probability  distribution  $\{  p(i)  \}$ defined  on  a  set  of
microstates $\Omega = \{ i \}$ \cite{discret} is given by
\begin{equation}
{\cal  S}_\Omega \left  ( \{p(i)\}  \right  ) =  -\sum_{i \in  \Omega}
[p(i)]^q \ln_q p(i),
\label{tsallis}
\end{equation}
where the parameter $q$, determining the degree of non-extensivity, is
positive  in  order  to  ensure  the concavity  of  ${\cal  S}$.   The
$q-$logarithmic function is defined as: $\ln_q f = (f^{1-q}-1)/(1-q)$.
In  the  $q  \rightarrow   1$  limit  Eq.   \ref{tsallis}  reduces  to
Gibbs-Shannon's entropy ${\cal  S} = -\sum_i p(i) \ln  p(i)$. In order
to simplify the notation, we  will indicate the set $\Omega$ only when
necessary. The  novelty of the statistical  entropy (\ref{tsallis}) is
that it does not satisfy additivity.  Instead, for two systems A and B
described by independent probability distributions:
\begin{equation}
{\cal S}(A+B)={\cal S}(A)+{\cal S}(B)+ (1-q){\cal S}(A){\cal S}(B). 
\label{nonadditivity}
\end{equation} 
For a  given physical system the  equilibrium probability distribution
$\{p^*(i)\}$ corresponds  to the distribution that  maximizes $\cal S$
under the normalization condition  ($\sum_{i \in \Omega} p(i) =1$) and
the  relevant   constraints  imposed  by   the  available  statistical
information on  the system.  The thermodynamic  equilibrium entropy is
then identified with $k {\cal S}^*$, where ${\cal S}^*= {\cal S} \left
( \{p^*(i)\}\right )$ and $k$  is the Boltzmann constant.  In the case
of an  isolated (microcanonical) system no  statistical information is
available   and   maximization  of   (\ref{tsallis})   leads  to   the
equiprobability  distribution on  $\Omega$.   Difficulties arise  when
trying  to study  non-isolated  systems for  which  some constants  of
motion $X_{\alpha}(i)$ in the isolated system, such as the energy, the
volume,  the magnetization,  the  number of  particles etc...,  become
fixed  only  on  average  ($\langle  X_\alpha  \rangle$).   The  index
$\alpha$ extends over the number of such controlled observables of the
non-isolated system.   It is worth  noting that a crucial  property of
such  observables,  required  by  the  foundations  of  thermodynamics
\cite{Giles64}, is  that they are  additive, i.e. for  two independent
systems A and B:
\begin{equation}
\langle   X_{\alpha}(A+B)\rangle  =  \langle   X_{\alpha}(A)\rangle  +
\langle X_{\alpha}(B)\rangle
\end{equation}
Within  the  Jaynes  scheme,  imposing knowledge  of  the  statistical
information  (i.e.  the  controlled average  values  $\langle X_\alpha
\rangle$)  on  the  maximization  procedure, leads  to  the  standard,
canonical,  isobaric,  isofield,  grand-canonical, etc...   ensembles.
Two different strategies can be adopted when using Tsallis statistical
entropy \cite{Tsallis98}:

(i) Unbiased averaging: when the average quantities of the microscopic
properties are defined in the standard way $\langle X_{\alpha} \rangle
= \sum_{i\in  \Omega} p(i)  X_{\alpha}(i)$, the Legendre  structure of
standard  thermodynamics is not  recovered \cite{Tsallis98},  i.e.  in
equilibrium, the  partition function cannot  be related to  a Legendre
transformation of  the entropy.  For this reason  this scheme  has not
been studied very much.

(ii) Biased  averaging: A Legendre structure is  recovered if averages
are defined  as: $\langle X_{\alpha}  \rangle_q = \sum_{i  \in \Omega}
P_q(i)  X_{\alpha}(i)$, where  the  averages are  performed using  the
so-called q-scort  probabilities: $P_q(i)=[p(i)]^q/\sum_{i \in \Omega}
[p(i)]^q$.  After maximization when  this choice is used, one recovers
a fundamental thermodynamic identity which reads:
\begin{equation}
d  {\cal   S}^*  =  \sum_{\alpha}  y_{\alpha}   d  \langle  X_{\alpha}
\rangle_q,
\label{termo}
\end{equation}
where $y_{\alpha}$ correspond to the Lagrange parameters necessary for
keeping  the average constraints  $\langle X_{\alpha}  \rangle_q$ when
performing the  maximization of the statistical  entropy. The physical
meaning  of   the  parameters  $y_{\alpha}$  is,   at  present,  still
controversial , and they are not necessarily those that control mutual
equilibrium  between   thermodynamic  systems  (physical  temperature,
pressure, magnetic  field, ...). Besides,  the equilibrium statistical
entropy can be expressed, in  any statistical ensemble, as ${\cal S^*}
= \ln_q Z_q$, where $Z_q$ plays the role of the partition function and
can be written as:
\begin{equation}
Z_q = \sum_{i \in \Omega} e_q^{-\sum_{\alpha} y_{\alpha} F(X_{\alpha}(i))}     
\end{equation}
with $F(X)=(X-\langle X  \rangle_q)/\sum_{i\in \Omega} [p(i)]^q$.  The
$q$-exponential function is the inverse of $\ln_q$ and is defined as :
$e_q^f =  [1 +  (1-q)f]^{\frac{1}{1-q}}$.  This definition  only makes
sense for $1 + (1-q)f>0$ and, in practice, this means that within this
biased  scheme only  the range  $0< q\leq  1$ has  a  physical meaning
\cite{Abe2001,Toral2001}.

It is worth  noting that a third "averaging"  procedure, consisting of
using the above proposed  biased averaging, but without normalization,
has  been  shown \cite{Tsallis98}  to  be  equivalent (after  suitable
redefinition of  the Lagrange  multipliers) to this  normalized biased
averaging strategy.

At this  point it is convenient  to analyze the  scaling properties of
the  non-extensive  statistical  entropy  ${\cal  S}$  satisfying  the
property  (\ref{nonadditivity}).   By  recursively  applying  equation
(\ref{nonadditivity}) to $\lambda$ identical systems one obtains:
\renewcommand{\arraystretch}{0.5}
\arraycolsep=0.1mm
\begin{eqnarray}
\nonumber
{\cal  S}(\lambda   A)  & = &   \left  (  \begin{array}{c}   \lambda  \\  1
\end{array} \right ) {\cal S}(A) + \left ( \begin{array}{c} \lambda \\
2 \end{array}  \right ) (1-q)  {\cal S}(A)^2+ \cdots  \\ & &  \cdots +
\left (
\begin{array}{c}    \lambda    \\    \lambda    \end{array}\right    )
(1-q)^{\lambda-1} {\cal S}(A)^{\lambda},
\end{eqnarray}
which can be easily expressed in the compact form:
\begin{equation}
{\cal  S}(\lambda  A)  =  \frac{1}{1-q}  \left \{  [  1  +  (1-q){\cal
S}(A)]^\lambda -1 \right\}.
\label{Newton}
\end{equation}
Of  course,  for $q\rightarrow  1$  the  usual  scaling for  extensive
systems is recovered.  

Let  us now  concentrate on  the analysis  of the  biased  scheme, and
consider  the  statistical entropy  ${\cal  S}^*$  of the  equilibrium
states  as a  function of  the  set of  extensive quantities  $\langle
X_{\alpha}   \rangle_q$.   Under   the   assumption  that   expression
(\ref{nonadditivity}) [and therefore (\ref{Newton})] can be applied to
the  equilibrium  distributions  and  thus  to ${\cal  S}^*$,  we  can
substitute  ${\cal  S}^*(A)$  by  ${\cal  S}^*(\{  \langle  X_{\alpha}
\rangle_q  \})$.  Differentiating  the expression  (\ref{Newton}) with
respect to $\lambda$ and taking $\lambda=1$ we obtain:
\begin{equation}
\sum_{\alpha} \langle X_{\alpha}  \rangle_q y_{\alpha} = \frac{1}{1-q}
\left [ 1+ (1-q)  {\cal S}^* \right ] \ln \left [  1+ (1-q) {\cal S}^*
\right ].
\label{Euler}
\end{equation}
Equation (\ref{Euler}) represents  a generalization of Euler's theorem
for   the   Tsallis   non-extensive   entropy.    By   differentiating
(\ref{Euler}) and  using the thermodynamic  identity (\ref{termo}), we
deduce that:
\begin{equation}
\sum_{\alpha} \langle  X_{\alpha} \rangle_q dy_{\alpha} =  \ln \left [
1+ (1-q) {\cal S}^* \right ] d{\cal S}^*
\label{GibbsDuhem}
\end{equation}
which  corresponds  to  the  Gibbs-Duhem  equation  for  non-extensive
systems. It  indicates that  the Lagrange parameters  $y_{\alpha}$ are
not intensive variables except for $q=1$.

The  equation obtained  suggests  a transformation  of variables  that
keeps  the  structure of  the  thermodynamic  identity  and allows  to
recover the standard Gibbs-Duhem equation.  Indeed, defining:
\begin{equation}
{\hat {\cal  S}^*} =  \int \frac{d {\cal  S}^*}{1+(1-q) {\cal  S}^*} =
\frac{\ln \left [ 1+ (1-q) {\cal S}^* \right ]}{1-q}
\label{Reny}
\end{equation}
and 
\begin{equation}
{\hat y}_{\alpha} = \frac{y_{\alpha}}{1+ (1-q) {\cal S}^*},
\label{prima}
\end{equation}
expressions (\ref{termo})  and (\ref{Euler})  can be rewritten  in the
form:
\begin{equation}
d{\hat   {\cal  S}^*}=\sum_{\alpha}   {\hat   y_{\alpha}}  d   \langle
X_{\alpha} \rangle_q,
\end{equation}
\begin{equation}
\sum_{\alpha} \langle X_{\alpha} \rangle_q d {\hat y_{\alpha}} =0.
\end{equation}
Therefore,  $\hat  {\cal S^*}$  is  an  extensive statistical  entropy
(satisfying  ${\hat  {\cal  S}^*}(\lambda  A) =  \lambda  {\hat  {\cal
S}^*}(A)$)  and  ${\hat  y_{\alpha}}$  are  the  conjugated  intensive
variables of $\langle X_{\alpha}  \rangle_q$.  It should also be noted
that the condition  $q\leq 1$ (necessary for the  transformation to be
well defined) ensures that ${\cal S}^*$ and $\hat {\cal S}^*$ have the
same concavity (thus the same maxima and minima).

The above  transformation of variables  (\ref{Reny}) and (\ref{prima})
has partially been suggested by different authors during recent years.
Tsallis  has already  shown the  extensive properties  of  $\hat {\cal
S}^*$ \cite{Tsallis88}, which in  fact had been proposed several years
before  by R\'enyi  \cite{Renyi70}.   The intensive  character of  the
variables  $\hat  y_{\alpha}$  was  recently  pointed out  by  Abe  et
al. \cite{Abe2001}, who showed  that such variables control the mutual
equilibrium  between thermodynamic  systems.  Finally,  the  fact that
standard thermodynamics is fully recovered after the transformation of
variables proposed  above has also been implicitly  suggested by Toral
\cite{Toral2001} within the framework of the microcanonical ensemble.

Another aspect which  is still under debate refers  to the calculation
of   statistical   fluctuations   within  Tsallis   statistics.    The
computation  of   the  variances  of   extensive  quantities  $\langle
X_{\alpha}^2     \rangle$    has     not    been     clearly    solved
\cite{Jedrzejewski2001} not only within the unbiased averaging scheme,
but  also  using  biased  averaging.   Here we  provide  a  method  of
computation  of fluctuations  which is  based on  a  generalization of
Einstein's  fluctuation formula  and is  independent of  the averaging
scheme.  Moreover,  for the case of  biased averaging, the  use of the
transformation   of   variables   proposed  above,   allows   standard
thermodynamic  fluctuations  to  be  recovered. The  method  has  been
recently proposed for the  study of fluctuations in standard extensive
statistics \cite{Caticha,Planes}.

In order  to compute fluctuations  within a very general  framework we
must  study  how  the  distribution  $p^*(i)$  changes  under  virtual
displacements.  Such displacements can  be understood as a consequence
of an  internal constraint  which causes the  deviation of  the system
from the equilibrium  state, or as a consequence of  a contact with an
external  bath  which allows  changes  in  parameters  which would  be
constant  under  total  isolation.    Let  us  consider  that  such  a
displacement  can  be characterized  by  a multidimensional  parameter
$\kappa$ which takes values on  a continuous set $\cal R$ and vanishes
at equilibrium.  We  define the set of functions $  \{ p^*(i | \kappa)
\}$  as the  displaced (conditional)  probability distributions  for a
given $\kappa$  obtained after maximization  of (\ref{tsallis}).  Note
that $p^*(i | 0 )= p^*(i)$.

The key point for characterizing fluctuations is to consider the joint
probability distribution $p(i , \kappa  )$ defined on the enlarged set
$\Omega  \times  {\cal R}$.  This  means  that  we are  assuming  that
$\kappa$ is a fluctuating variable and that the system is described by
simultaneously specifying  $i$ and $\kappa$.  The problem  now is to
determine $p(i,\kappa)$. Using the same philosophy used for
the determination  of equilibrium  distributions $p^*(i|\kappa)$, we
propose to generalize the maximum statistical entropy principle to the
determination of $p^*(i,\kappa)$ on $\Omega \times \cal R$. This means
maximizing:
\begin{equation}
{\cal S}_{\Omega \times  {\cal R}} \left ( \{p(i,\kappa)\}  \right ) =
-\int_{\cal R} \sum_{i \in  \Omega} [p(i,\kappa)]^q \ln_q p(i, \kappa) d
\kappa,
\label{tsallis2}
\end{equation}
where we have  assumed that the measure of the  lack of information on
$\Omega   \times  \cal   R$   is  also   given   by  Tsallis   entropy
\cite{general}.

Taking  into account  the mathematical  definition of  the conditional
probability  $p(i,\kappa)  = p(\kappa)  p^*(i|\kappa)$  and using  the
standard property $\ln_q f g = \ln_q f + \ln_q g + (1-q) \ln_q f \ln_q
g$   and  the   normalization  condition   for   $p^*(i|k)$,  equation
(\ref{tsallis2}) can be written as:
\begin{equation}
{\cal S}_{\Omega \times {\cal  R}} = \int_{\cal R} [p(\kappa)]^q {\cal
 S}_{\Omega}^*(\kappa)+{\cal S}_{\cal R}(\{ p(\kappa) \}),
\label{tsallis3}
\end{equation}
where   ${\cal  S}_{\Omega}^*(\kappa)={\cal   S}_{\Omega}\left   (  \{
p^*(i|\kappa)  \} \right  )$ is  nothing more  than  the thermodynamic
entropy of the virtually  displaced system.  Therefore, the problem of
determining   the  distribution   of  fluctuations   reduces   to  the
maximization  of  the  functional  (\ref{tsallis3})  with  respect  to
$p(\kappa)$ under the normalization condition $\int_{\cal R} p(\kappa)
d \kappa = 1$.  This yields:
\begin{equation}
p^*(\kappa) = \frac{e_q^{{\cal S}^*(\kappa)}} {\int_{\cal R}e_q^{{\cal
S}^*(\kappa)} d \kappa}.
\label{einstein}
\end{equation}
Since  the  $e_q$  function  converges  to  the  standard  exponential
function  in the  limit $q  \rightarrow  1$, the  above equation  just
reduces  to the well-known  Einstein formula  in the  extensive limit.
This  probability  distribution  for  fluctuations  $p^*(\kappa)$  is,
therefore, the most probable (or the least biased) distribution of the
fluctuating    parameter   $\kappa$.     Also   note    that   formula
(\ref{einstein}) can be deduced in a non-rigorous way (as done in many
statistical mechanics text-books in  the case of extensive statistics)
by   inversion  of   the   generalized  Boltzmann   formula  for   the
microcanonical  ensemble: ${\cal  S}^*  =\ln_q W$,  where  $W$ is  the
number of states in $\Omega$.
 
Taking into account  the fact that ${\cal S}^*(\kappa)$  is maximum at
$\kappa=0$ we can expand it in powers of $\kappa$, as:
\begin{equation}
{\cal    S}^*(\kappa)   =    {\cal    S}^*(0)+\frac{1}{2}   \left    (
\frac{\partial^2  {\cal S}^*}{\partial \kappa^2}  \right )  \kappa^2 +
\dots
\end{equation}
with  $ \left  (  \partial^2  {\cal S}^*  /  \partial \kappa^2  \right
)<0$. Substituting  in (\ref{einstein}),  we find that  small $\kappa$
fluctuations are Gaussian distributed according to:
\begin{equation}
p^*(\kappa)  \sim  \exp  \left  \{  - \frac{1}{2}  \;  \frac{-\left  (
\frac{\partial^2  {\cal  S}^*}{\partial  \kappa^2}\right  )}{1+  (1-q)
{\cal S}^*(0)} \kappa^2 \right \} .
\end{equation}
This  formula, as  well  as (\ref{einstein}),  is  independent of  the
different  averaging  procedures explained  in  the introduction.   It
shows that the probability distribution for fluctuations of measurable
quantities depends on  $q$ as well as on the  entropy of the reference
equilibrium state around which fluctuations take place.

In order to  compute the variance $\langle \kappa^2  \rangle $ the two
averaging schemes discussed above  can be considered. Nevertheless, if
$\kappa$  is a macroscopic  quantity, the  unbiased scheme  seems more
appropriate.  The two choices yield  very similar results in the limit
of small fluctuations.  If we  use unbiased averaging, the variance is
given by:
\begin{equation}
\langle \kappa^2 \rangle = -  \left [ ( 1+ (1-q) {\cal S}^*(0) \right ]
\left ( \frac{\partial^2 {\cal S}^*}{\partial \kappa^2} \right )^{-1},
\label{variancia}
\end{equation}
whereas in  the case when we  use the biased  averaging procedure, one
gets  the  same  result,  but  it  is  divided  by  $q$  (the  q-scort
probability of a Gaussian distribution  is also Gaussian). It is worth
noting that within the  two interpretations, as $q<1$ fluctuations are
larger than those corresponding to the extensive case $q=1$.

It is  also interesting to  note that the transformation  of variables
(\ref{Reny}) and (\ref{prima}) suggested  in the previous section 
allows standard  thermodynamics to be recovered within the  biased scheme,
and can be  used to  rewrite equation (\ref{einstein})  in terms  of $\hat
{\cal S}^*$.   Using the  fact that $\cal  S^*$ and $\hat  {\cal S}^*$
have the same concavity, a straightforward calculation yields:
\begin{equation}
p^*(\kappa)  = \frac{e ^  {\hat {\cal  S}^*(\kappa)}}{\int_{\cal R}e ^
{\hat {\cal S}^*(\kappa)} d \kappa}
\label{einstein2}
\end{equation}
which is the standard Einstein formula, and the variance transforms
into:
\begin{equation}
\langle  \kappa^2  \rangle  =  \left  (  \frac{\partial^2  \hat  {\cal
S}^*}{\partial \kappa^2} \right )^{-1}.
\label{variancia2}
\end{equation}
Therefore the  transformation of variables allows not only
standard thermodynamics to be recovered, but also the same statistical fluctuations.
For  instance,  a  simple  calculation for  the  energy  fluctuations
$\Delta E$ from this expression gives:
\begin{equation}
\langle \Delta E  ^2 \rangle = - \left (  \frac{\partial E }{ \partial
\hat{ \beta}}\right ),
\end{equation}
where  $\hat  \beta=  \beta/[1+(1-q)  {\cal S}^*]$  is  the  intensive
physical inverse  temperature, and  $\beta$ is the  Lagrange parameter
associated with $\langle E \rangle_q$.

In  summary, we  propose that  Tsallis statistics,  within  the biased
averaging scheme, can be mapped into standard thermodynamics.  This is
done by  a suitable transformation  of variables, which  appears after
the derivation of the  generalized Gibbs-Duhem equation resulting from
the scaling properties of  Tsallis entropy.  The intensive generalized
forces in  such framework arise  naturally from the  transformation of
variables.      In      addition,     the     standard     equilibrium
fluctuation-dissipation  theorem is  also  recovered.  This  indicates
that, besides the  thermodynamic equivalence, statistical fluctuations
also  behave according  to standard  statistical  mechanics. Moreover,
this ensures the equivalence of the equilibrium response functions.


This  work has  received  financial support  from  the CICyT  (Spain),
project   MAT98-0315   and  from   the   CIRIT  (Catalonia),   project
2000GR00025.  We acknowledge  fruitful discussions with our colleagues
J.Casademunt, T.Cast\'an, Ll.Ma\~nosa, J.Ort\'{\i}n and J.M.Sancho.


\begin{references}
\bibitem{discret}  For   the  sake  of  simplicity   and without  loss  of
generality we consider $\Omega$ as a discrete set of states.
\bibitem{general} A generalization of such an assumption could be done by
assuming that the non-extensivity parameter $q$ depends on $\kappa$.
\end{references}
\end{document}